\documentclass[10pt]{article}
\usepackage{multicol}
\usepackage{graphicx}
\usepackage{amsmath}
\usepackage[a4paper]{geometry}
\usepackage{hyperref}
\usepackage{rotating}

\begin{document}

\begin{center}
{\Large \bf Study of proton, deuteron and triton at 54.4 GeV}

\vskip1.0cm

M.~Waqas$^{1,}${\footnote{Corresponding author. Email (M.Waqas):
waqas\_phy313@yahoo.com; waqas\_phy313@ucas.ac.cn}},G. X. Peng$^{1,2,3}$ {\footnote{Corresponding author. Email (G. X. Peng): gxpeng@ucas.ac.cn}}
\\

{\small\it  $^1$ School of Nuclear Science and Technology, University of Chinese Academy of Sciences,
Beijing 100049, China,

$^2$ Theoretical Physics Center for Science Facilities, Institute of High Energy Physics, Beijing 100049, China,

$^3$ Synergetic Innovation Center for Quantum Effects \& Applications, Hunan Normal University, Changsha 410081, China}

\end{center}

\vskip1.0cm

{\bf Abstract:} Transverse momentum spectra of proton, deuteron
and triton in gold-gold (Au-Au) collisions at 54.4 GeV are
analyzed in different centrality bins by the blast wave model with
Tsallis statistics. The model results are approximately in
agreement with the experimental data measured by STAR
Collaboration in special transverse momentum ranges. We extracted
the kinetic freeze out temperature, transverse flow velocity and
freeze out volume from the transverse momentum spectra of the
particles. It is observed that the kinetic freeze out temperature
is increasing from central to peripheral collisions. However the
transverse flow velocity and freeze out volume decrease from
central to peripheral collisions. The present work reveals the
mass dependent kinetic freeze out and volume differential freeze
out scenario in collisions at STAR Collaboration. In addition, the
parameter q characterizes the degree of non-equilibrium of the
produced system, and it increase from central to peripheral
collisions and increase with mass.
\\



\vskip1.0cm

\begin{multicols}{2}

{\section{Introduction}} The two important stages in the evolution
system are chemical and kinetic freeze out. The degree of
excitation of the interacting system at the two stages are
different from each other. The chemical and kinetic freeze out
temperatures are used to describe the different excitation degree
of two stages. In general, the ratios of different kinds of
particles are no longer change at the stage of chemical freeze
out. The chemical freeze out temperature can be obtained from
different particle ratios in the framework of thermal model
[1--3]. The transverse momentum spectra of different particles are
no longer changed at the stage of kinetic freeze out and
thermal/kinetic freeze out temperature can be obtained from the
transverse momentum spectra according to hydrodynamical model [4].

It is important to point out that the transverse momentum spectra
even though in a narrow $p_T$ range, but it contains both the
contribution of thermal motion and transverse flow velocity. The
random thermal motion reflect the excitation and the transverse
flow velocity  reflects the degree of expansion of interacting
system. In order to extract the kinetic freeze out temperature
($T_0$), we have excluded the contribution of transverse flow
velocity ($\beta_T$), that is to disengage the random thermal
motion and transverse flow velocity. There are various methods to
disengage the two issues. The methods include but are not limited
to blast wave fit with boltzmann Gibbs statics [5--7], blast wave
model with Tsallis statistics [8--10], and alternative methods
[11--17].

The dependence of $T_0$ and $\beta_T$ on centrality is very
complex situation. There are two schools of thought.  (1) $T_0$
increase decrease from central to peripheral collisions [18--21]
(2) $T_0$ increase from central to peripheral collision [22--23].
Both have their own explanations. Larger $T_0$ in the central
collisions explain higher degree of excitation of the system due
to more violent collisions, while smaller $T_0$ in the central
collisions indicates longer liver fireball in the central
collisions. It is very important to find out which collision
system contains larger $T_0$. Furthermore, there are several
opinions about the freeze out of particles which include single,
double or multiple kinetic freeze out. It is also very important
to dig our the correct freeze out scenario.

In the present work, we will analyze the $p_T$ spectra  of proton,
deuteron and triton and will extract $T_0$ and $\beta_T$. Deuteron
and triton are light nuclei. The fundamental mechanism, for light
nuclei production in relativistic heavy ion collision is not well
understood [24--26]. Coalescence of anti-(nucleons) is possible
approach [27--31]. Because of small binding energies (d with 2.2
MeV and t with 8.8 MeV), the light nuclei cannot persists when the
temperature is much higher than their binding energy. The typical
kinetic freeze out temperature is around 100 MeV for light
hadrons, so they might disintegrate and be formed again by final
state coalescence after nucleons are decoupled from the hot and
dense system. Hence the study of the light nuclei can be useful in
the extraction of information of nucleons distribution at the
freeze out [27, 30, 32].

Before going to the formalism, we would point out the concept of
volume is important in high energy collisions. The volume occupied
by the ejectiles when the mutual interactions become negligible,
and the only force they feel is the columbic repulsive force, is
called the kinetic freeze-out volume (V ). Various freeze-out
volumes occurs at various freeze-out stages, but we are only
focusing on the kinetic freeze-out volume $V$ in the present work.
The information about the information of the co-existence of
phase-transition, and is important in the extraction of
multiplicity, micro-canonical heat capacity and it's negative
branch or shape of the caloric curves under the thermal
constraints can be obtained $V$.

The remainder of the paper consists of method and formalism in
section 2, followed by the results and discussion in section 3. In
section 4, we summarized our main observations and conclusions.
\\

{\section{The method and formalism}} In high energy collisions
there are two types particles production process. (1) soft process
ad (2) hard process. For soft process, there are various methods
which includes but are not limited to blast wave model with
boltzmann Gibbs statistics [5--7], blast wave model with Tsallis
statistics [8--10], Hagedorn thermal model [20] and Standard
distribution [33, 34] etc. We are interested in blast wave model
with Tsallis statistics. According to [8], the blast wave fit with
Tsallis statistics results in the probability density function be
\begin{align}
f_1(p_T)=&\frac{1}{N}\frac{\mathrm{d}N}{\mathrm{d}p_\mathrm{T}}=C \frac{gV}{(2\pi)^2} p_T m_T \int_{-\pi}^\pi d\phi\int_0^R rdr \nonumber\\
& \times\bigg\{{1+\frac{q-1}{T_0}} \bigg[m_T  \cosh(\rho)-p_T \sinh(\rho) \nonumber\\
& \times\cos(\phi)\bigg]\bigg\}^\frac{-1}{(q-1)}
\end{align}
where $C$ denotes the normalization constant that leads the
integral in Eq. (1) to be normalized to 1, $g$ is the degeneracy
factor which is different for different particles based on
$g_n$=2$S_n$+1, $m_T=\sqrt{p_T^2+m_0^2}$ is the transverse mass,
$m_0$ is denotes rest mass of the particle, $\phi$ shows the
azimuthal angle, r is the radial coordinate, $R$ is the maximum
$r$, $q$ represents the measure of degree of deviation of the
system from an equilibrium state, $\rho=\tanh^{-1} [\beta(r)]$ is
the boost angle, $\beta(r)=\beta_S(r/R)^{n_0}$ is a self-similar
flow profile, $\beta_S$ represents the flow velocity on the
surface,as mean of $\beta(r)$,$\beta_T=(2/R^2)\int_0^R
r\beta(r)dr=2\beta_S/(n_0+2)=2\beta_S/3$, and $n_0$ =1.
Furthermore, the index $-1/(q-1)$ in Eq.(1) can be substituted by
$-q/(q-1)$ due to the reason that $q$ is being close to 1. This
substitution results in a small and negligible divergence in the
Tsallis distribution.

In case of a not too wide $p_T$ range, the above Eqn. can be used
to describe the $p_T$ spectra and we can extract $T_0$ and
$\beta_T$. But if we use the wide $p_T$ spectra, then the
contribution of hard scattering process can be considered.
According to quantum chromodynamics (QCD) calculus [35--37], the
contribution of hard process is parameterized to be an inverse
power law.
\begin{align}
f_H(p_T)=&\frac{1}{N}\frac{\mathrm{d}N}{\mathrm{d}p_\mathrm{T}}=
Ap_T \bigg( 1+\frac{p_T}{p_0} \bigg)^{-n},
\end{align}
which is the Hagedorn function [38, 39], $A$ is the normalization
constant while $p_0$ and $n$ are the free parameters.

The superposition of soft and hard scattering process can be used
if the $p_T$ spectra is distributed in a wide range. If Eqn. (1)
describes the contribution of soft process, then the contribution
of hard process can be described by Eqn. (2). To describe the
spectrum in a wide $p_T$ range, one can superpose the
two-component superposition like this
\begin{align}
f_0(p_T)=kf_S(p_T)+(1-k)f_H(p_T),
\end{align}
where $k$ denotes shows the contribution fraction of soft
excitation and  $(1-k)$ shows hard scattering process, $f_S$
denotes the soft process which contributes in the low $p_T$ region
and $f_H$ is the hard process which contributes in a whole $p_T$
region. The two contributions overlap each other in the low $p_T$
region.

We may also use the usual step function to superpose the two
functions. According to Hagedorn model [38]
\begin{align}
f_0(p_T)=A_1\theta(p_1-p_T) f_S(p_T) + A_2
\theta(p_T-p_1)f_H(p_T),
\end{align}
where $A_1$ and $A_2$ are the normalization constants that
synthesize $A_1$$f_S$$(p_1)$=$A_1$$f_H$$(p_1)$ and $\theta$(x) is
the usual step function.
\\

{\section{Results and discussion}} Fig. 1 presents the transverse
momentum ($p_T$) spectra [(1/2$\pi$$p_T$) $d^2$$N$/$dyd$$p_T$] of
proton, deuteron and triton in Au-Au collisions at
$\sqrt{s_{NN}}$=54.4 GeV. The $p_T$ spectra is distributed in
different centrality bins of  $0-10\%$, $10-20\%$, $20-40\%$,
$40-60\%$ and $60-80\%$ for $p$ and $d$, and $0-10\%$, $10-20\%$,
$20-40\%$ and $40-60\%$ for triton at $|y|<0.5$, where $|y|$
denotes the rapidity. The symbols represent the experimental data
measured by the STAR Collaborations [40] and the curves are our
fitting results by using the blast-wave model with Tsallis
statistics. Each panel is followed by its corresponding data/fit.
The related parameters, $\chi^2$ and degree of freedom (dof) are
listed in Table 1. One can see that Eq. (1) fits well the data in
Au-Au collisions at 54.4 GeV at the RHIC.
\begin{figure*}[htb!]
\begin{center}
\hskip-0.153cm
\includegraphics[width=15cm]{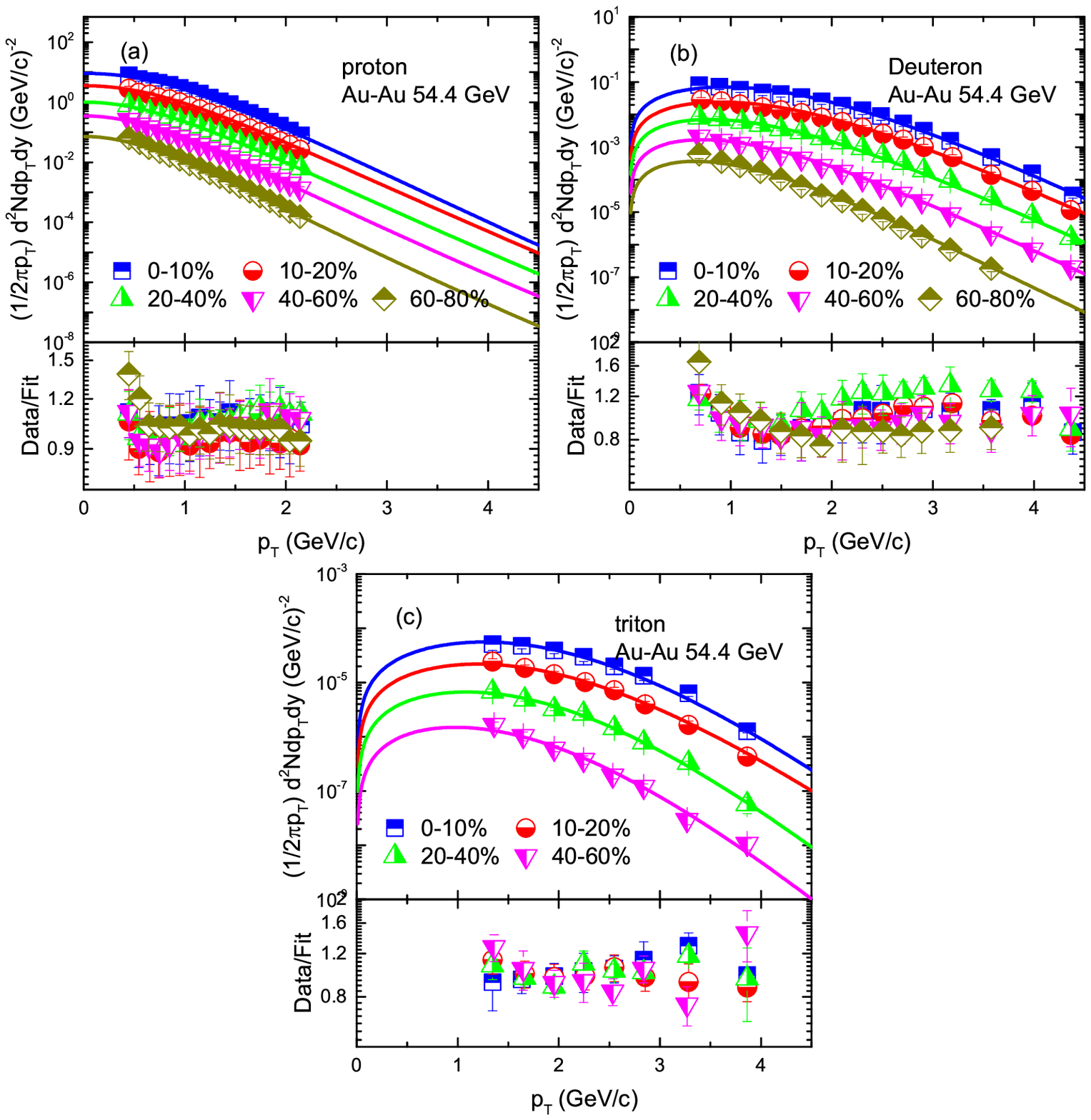}
\end{center}
Fig. 1. Transverse mass spectra of (a)-(c) $p$, $d$, and $t$
produced in different centrality bins in Au-Au collisions at
$\sqrt{s_{NN}}=54.4$ GeV. The symbols represent the experimental
data measured by the STAR Collaboration at $|y|<0.5$ [40]. The
curves are our fitted results by Eq. (1). Each panel is followed
by its corresponding ratios of Data/Fit.
\end{figure*}
\begin{figure*}[htb!]
\begin{center}
\hskip-0.153cm
\includegraphics[width=15cm]{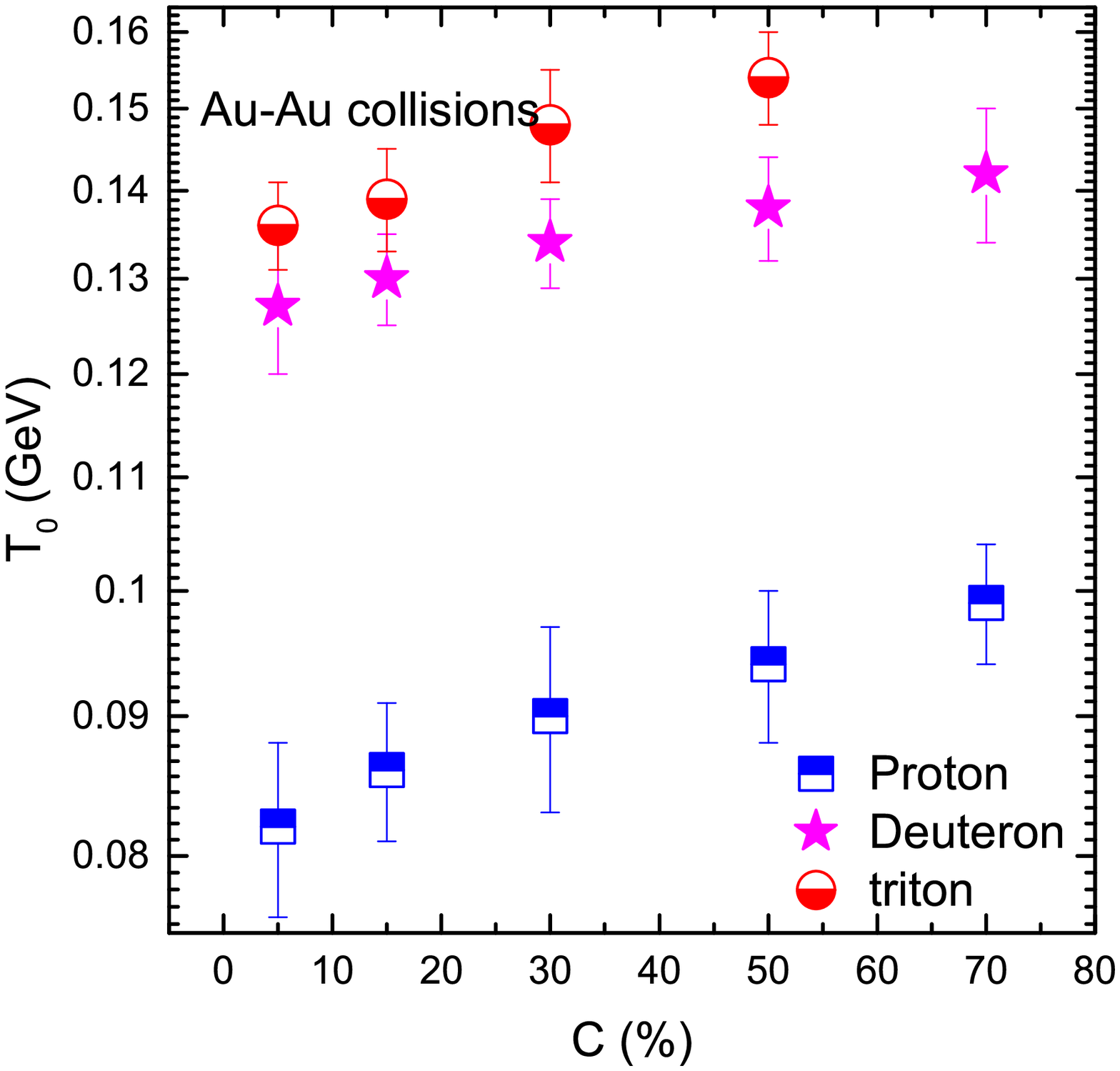}
\end{center}
Demonstrates the dependence of $T_0$ on centrality
\end{figure*}

To show the trend of the extracted parameters, Fig. 2 shows the
dependence of kinetic freeze out temperature on centrality. One
can see that $T_0$ in central collisions is smaller and it is
increasing with decrease of centrality which indicates a decrease
of lifetime of fireball from central collisions to peripheral
collisions. Furthermore, $T_0$ is observed to be mass dependent,
as it larger from triton, followed by deuteron and then proton
which means heavy particles freeze out early than lighter
particles.
\begin{figure*}[htb!]
\begin{center}
\hskip-0.153cm
\includegraphics[width=15cm]{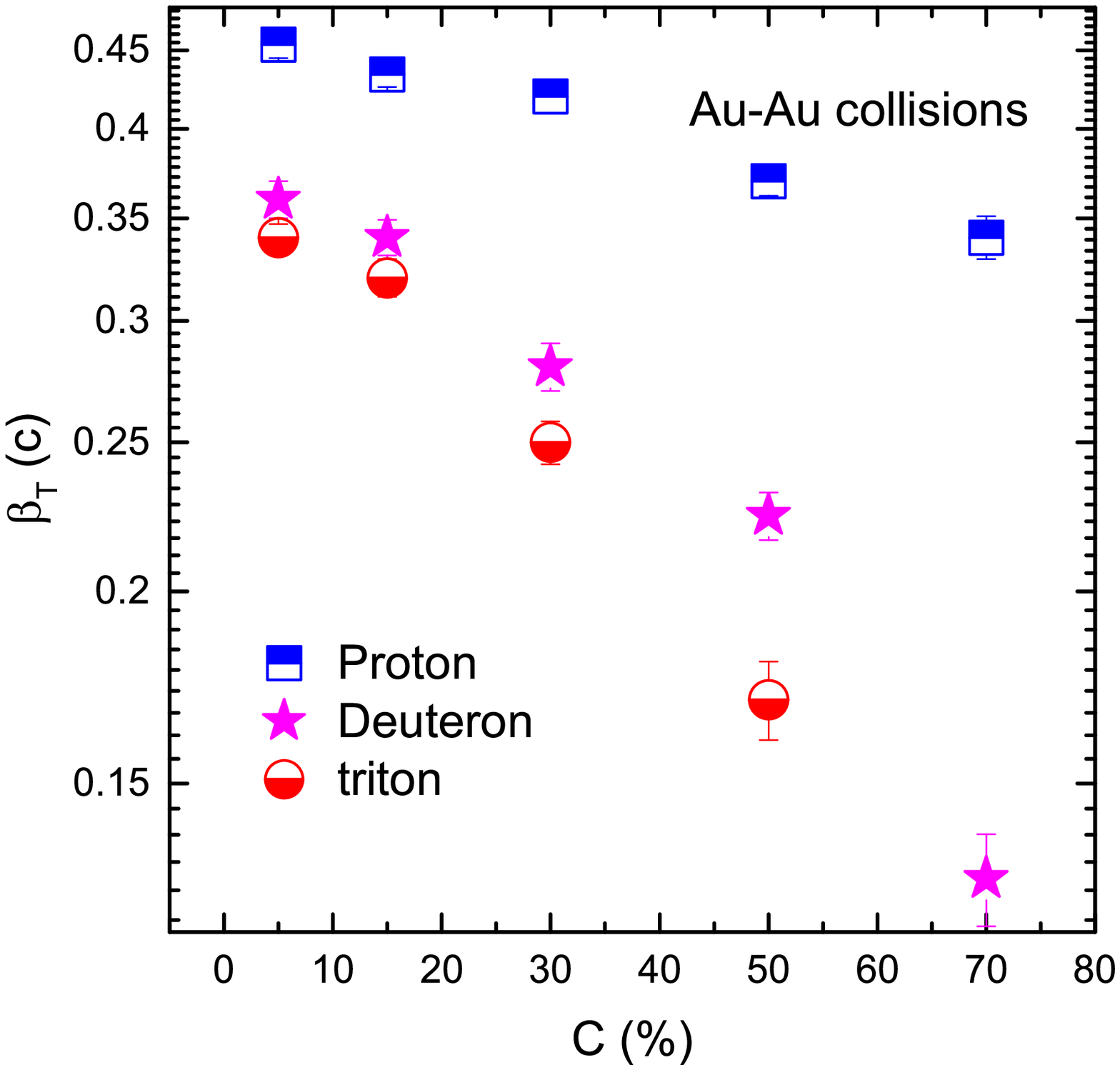}
\end{center}
Fig. 3. Demonstrates the dependence of $\beta_T$ on centrality.
\end{figure*}

Fig. 3 shows the centrality dependence of transverse flow
velocity. $\beta_T$ is observed to decrease with the decrease of
centrality due to the reason that in central collisions the system
undergoes more violent collisions and the system expands very
rapidly. In addition, $\beta_T$ is observed smaller for heavy
particles.
\begin{figure*}[htb!]
\begin{center}
\hskip-0.153cm
\includegraphics[width=15cm]{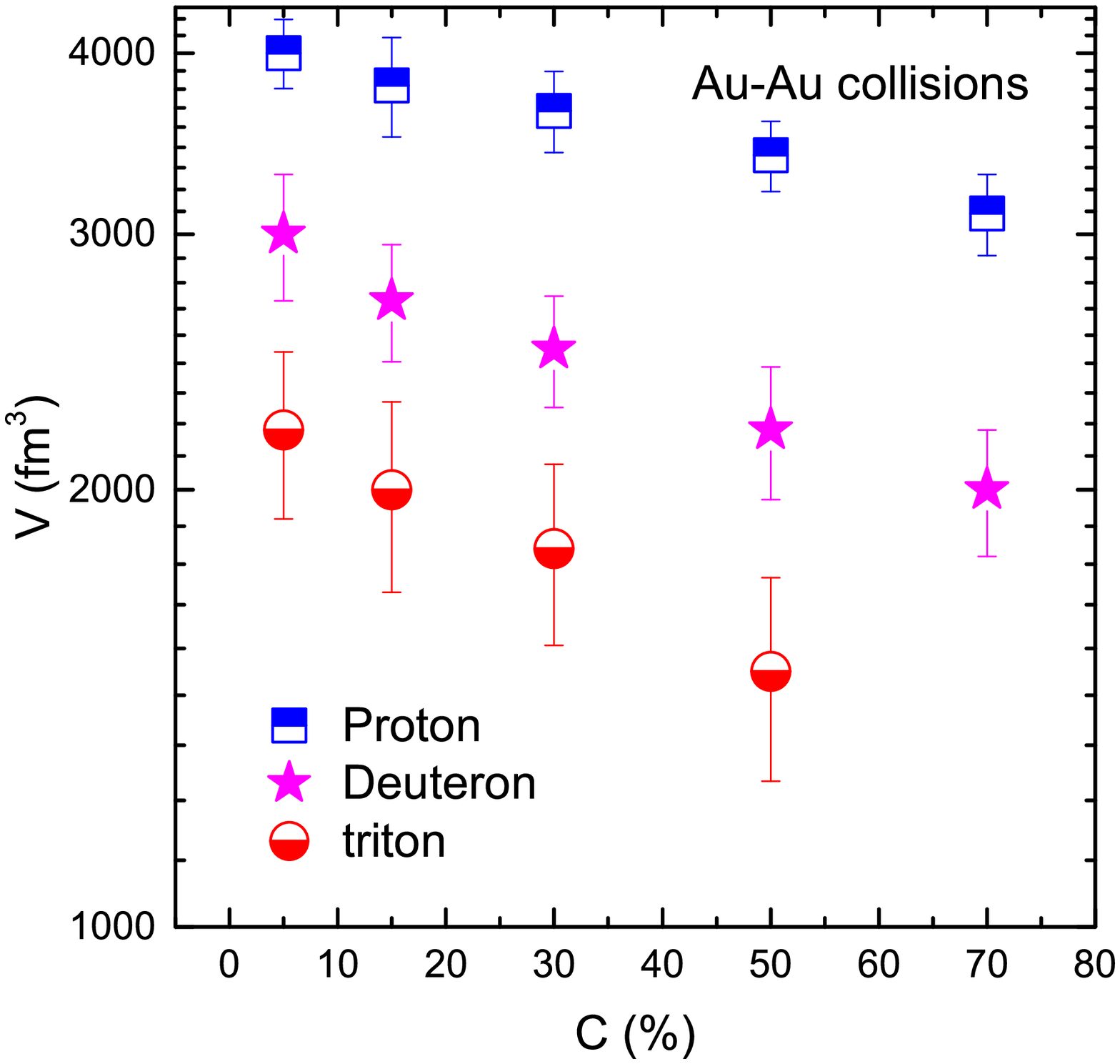}
\end{center}
Fig. 4. Demonstrates the dependence of $V$ on centrality.
\end{figure*}
\begin{figure*}[htb!]
\begin{center}
\hskip-0.153cm
\includegraphics[width=15cm]{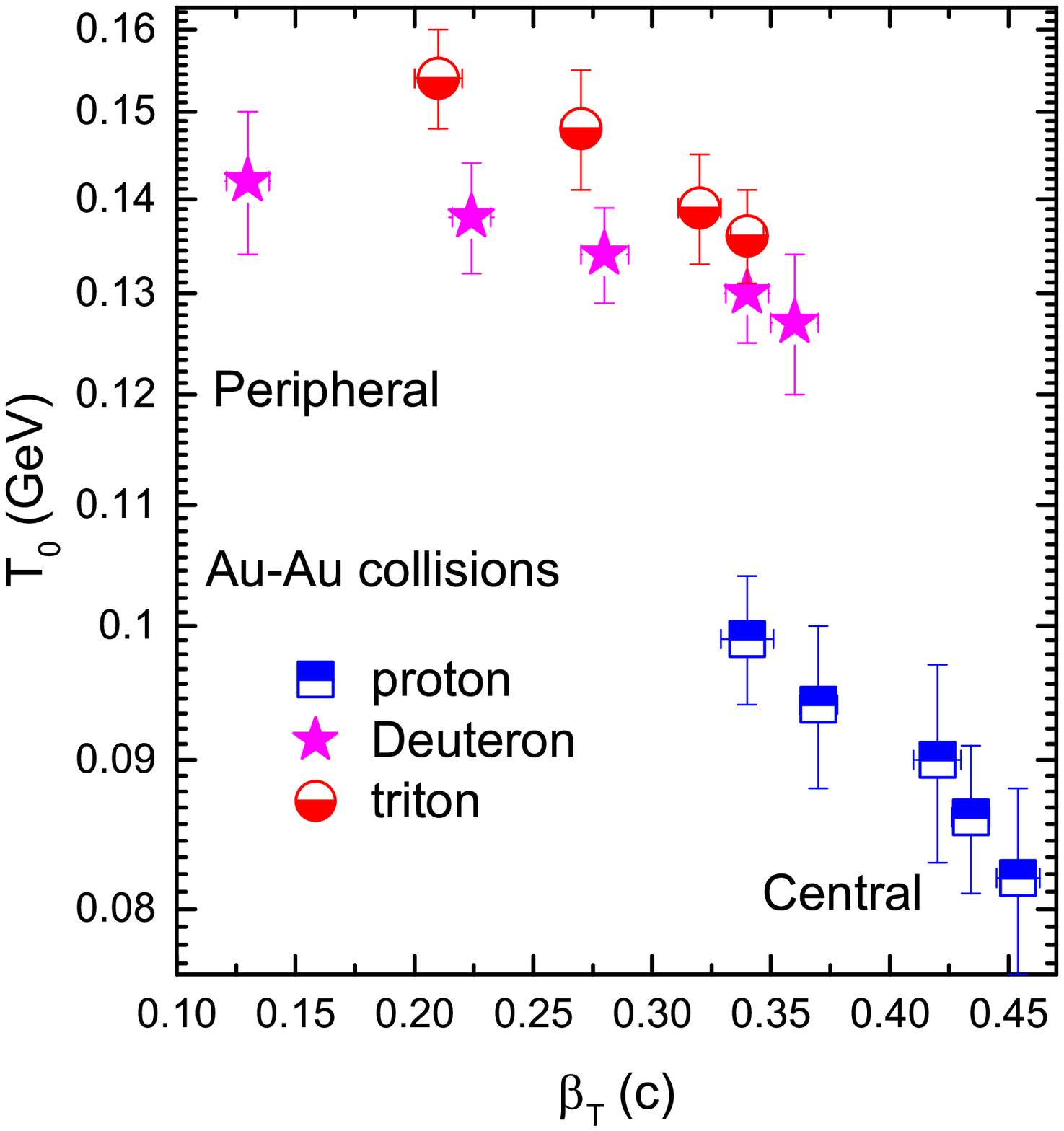}
\end{center}
Fig. 5. Variation of $T_0$ with $\beta_T$.
\end{figure*}
\begin{figure*}[htb!]
\begin{center}
\hskip-0.153cm
\includegraphics[width=15cm]{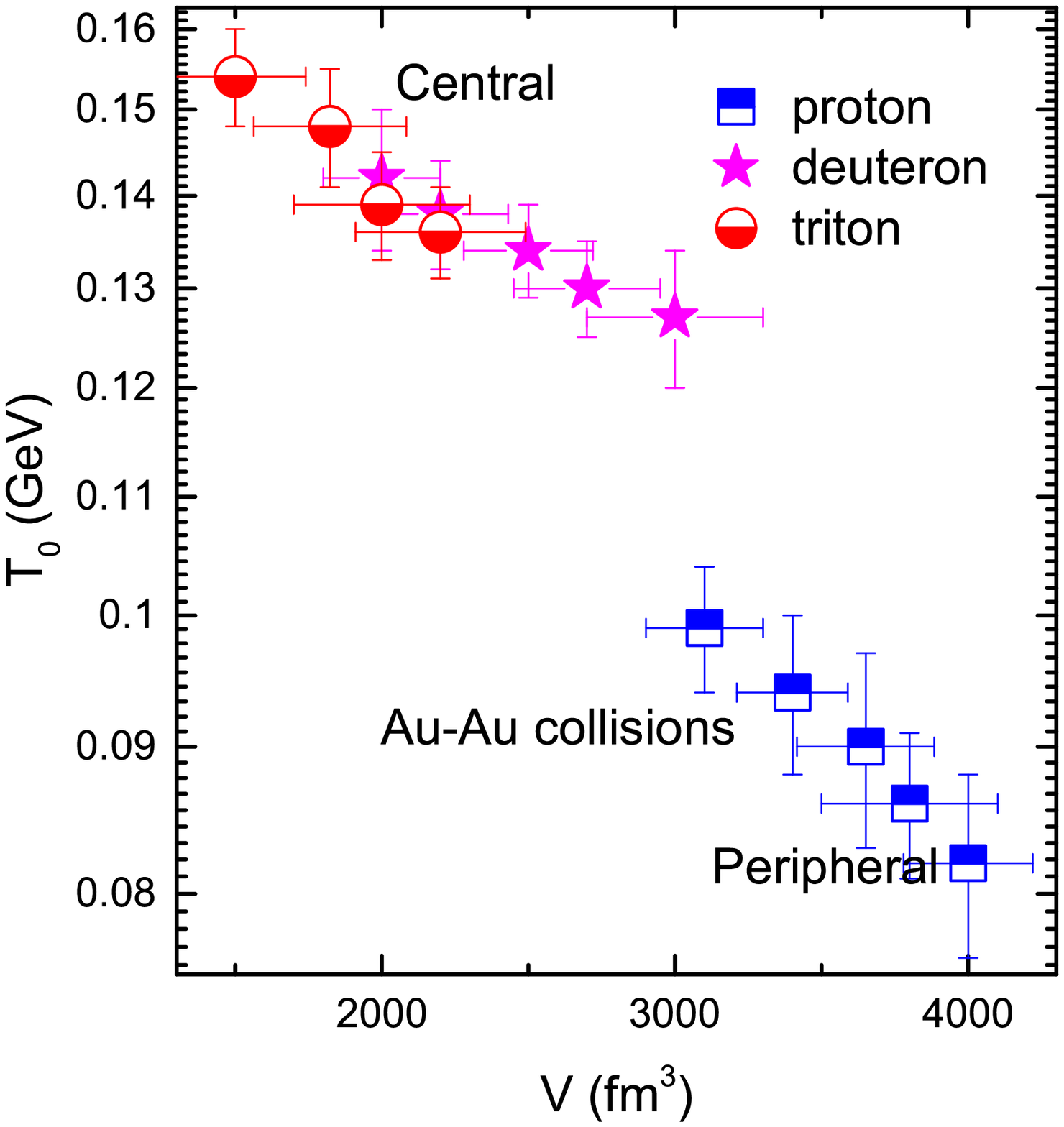}
\end{center}
Fig. 6. Variation of $T_0$ with $V$.
\end{figure*}
\begin{table*}
{\scriptsize Table 1. Values of free parameters $T_0$ and
$\beta_T$, V and q, normalization constant ($N_0$),
$\chi^2$, and degree of freedom (dof) corresponding to the curves
in Figs. 1--6. \vspace{-.50cm}
\begin{center}
\begin{tabular}{ccccccccc}\\ \hline\hline
Collisions       & Centrality       & Particle  & $T_0$ (GeV)   &
$\beta_T$ (c)    & $V (fm^3)$      & $q$          & $N_0$
& $\chi^2$/ dof \\ \hline
Fig. 1           & 0--10\%          & $p$       &$0.082\pm0.006$  & $0.454\pm0.009$  & $4000\pm220$ & $1.012\pm0.004$   &$0.019\pm0.005$                  & 5/13\\
Au-Au            & 10--20\%         & --        &$0.086\pm0.005$  & $0.434\pm0.008$  & $3800\pm300$ & $1.015\pm0.005$   &$0.0072\pm0.0004$                & 6/13\\
54.4 GeV          & 20--40\%         &--        &$0.090\pm0.007$  & $0.420\pm0.010$  & $3650\pm234$ & $1.018\pm0.007$   &$0.0019\pm0.0005$                & 11/13\\
                 & 40--60\%         & --        &$0.094\pm0.006$  & $0.370\pm0.008$  & $3400\pm190$ & $1.022\pm0.006$   &$6\times10^{-4}\pm6\times10^{-5}$& 7/13\\
                 & 60--80\%         & --        &$0.099\pm0.002$   & $0.340\pm0.011$   & $3100\pm200$ & $1.024\pm0.007$   &$1.2\times10^{-4}\pm4\times10^{-5}$& 29/13\\
\cline{2-8}
                 & 0--10\%          & $d$       &$0.127\pm0.007$  & $0.360\pm0.010$  & $3000\pm177$ & $1.004\pm0.006$   &$4.15\times10^{-4}\pm4\times10^{-5}$& 10/12\\
                 & 10--20\%         & --        &$0.130\pm0.005$  & $0.340\pm0.009$  & $2700\pm220$ & $1.007\pm0.005$   &$1.5\times10^{-4}\pm5\times10^{-5}$ & 11/12\\
                 & 20--40\%         &--         &$0.134\pm0.005$  & $0.280\pm0.010$  & $2500\pm180$ & $1.011\pm0.004$   &$5.1\times10^{-5}\pm6\times10^{-6}$ & 39/12\\
                 & 40--60\%         & --        &$0.138\pm0.006$  & $0.224\pm0.008$  & $2200\pm200$ & $1.013\pm0.005$   &$1.2\times10^{-5}\pm6\times10^{-6}$ & 3/12\\
                 & 60--80\%         & --        &$0.142\pm0.008$  & $0.130\pm0.009$  & $2000\pm200$ & $1.016\pm0.006$   &$2.7\times10^{-6}\pm4\times10^{-7}$ & 16/10\\
\cline{2-8}
                 & 0--10\%          & $t$    &$0.136\pm0.005$  & $0.340\pm0.007$  & $2200\pm165$ & $1.0002\pm0.004$   &$9.17\times10^{-7}\pm3\times10^{-8}$ & 4/3\\
                 & 10--20\%         & --     &$0.139\pm0.006$  & $0.320\pm0.009$  & $2000\pm150$ & $1.0006\pm0.006$  &$4\times10^{-7}\pm4\times10^{-8}$     & 3/3\\
                 & 20--40\%         &--      &$0.148\pm0.007$  & $0.250\pm0.008$  & $1823\pm135$ & $1.0009\pm0.007$  &$1.4\times10^{-7}\pm6\times10^{-8}$   & 12/8\\
                 & 40--60\%         & --     &$0.154\pm0.006$  & $0.170\pm0.010$  & $1500\pm180$ & $1.0012\pm0.006$ &$3.7\times10^{-8}\pm6\times10^{-9}$    & 6/3\\
\hline
\end{tabular}%
\end{center}}
\end{table*}

Fig. 4 is same as fig. 3 but shows the centrality dependent freeze
out volume. The freeze out volume decrease with decrease of
centrality due to decreasing the number of participant nucleons.
There are large number of binary collisions due to the
re-scattering of partons in central collisions and therefore the
system with more participants reaches quickly to equilibrium
state. Furthermore, the volume differential scenario is observed
and heavy particles is observed to have less freeze out volume and
this shows the early freeze out heavier particles. The different
freeze out of different particles exhibits different freeze out
surfaces different particles.

Fig. 5 and fig. 6 are the same. Fig. 5 shows the correlation of
$T_0$ and $\beta_T$  and fig. 6 show the correlation of $T_0$ and
$V$. one can see that both $T_0$ and $\beta_T$  and $T_0$ and $V$
exhibits a two dimensional anti-correlation band. Larger the
$T_0$, smaller the $\beta$ and $V$.

\begin{figure*}[htb!]
\begin{center}
\hskip-0.153cm
\includegraphics[width=15cm]{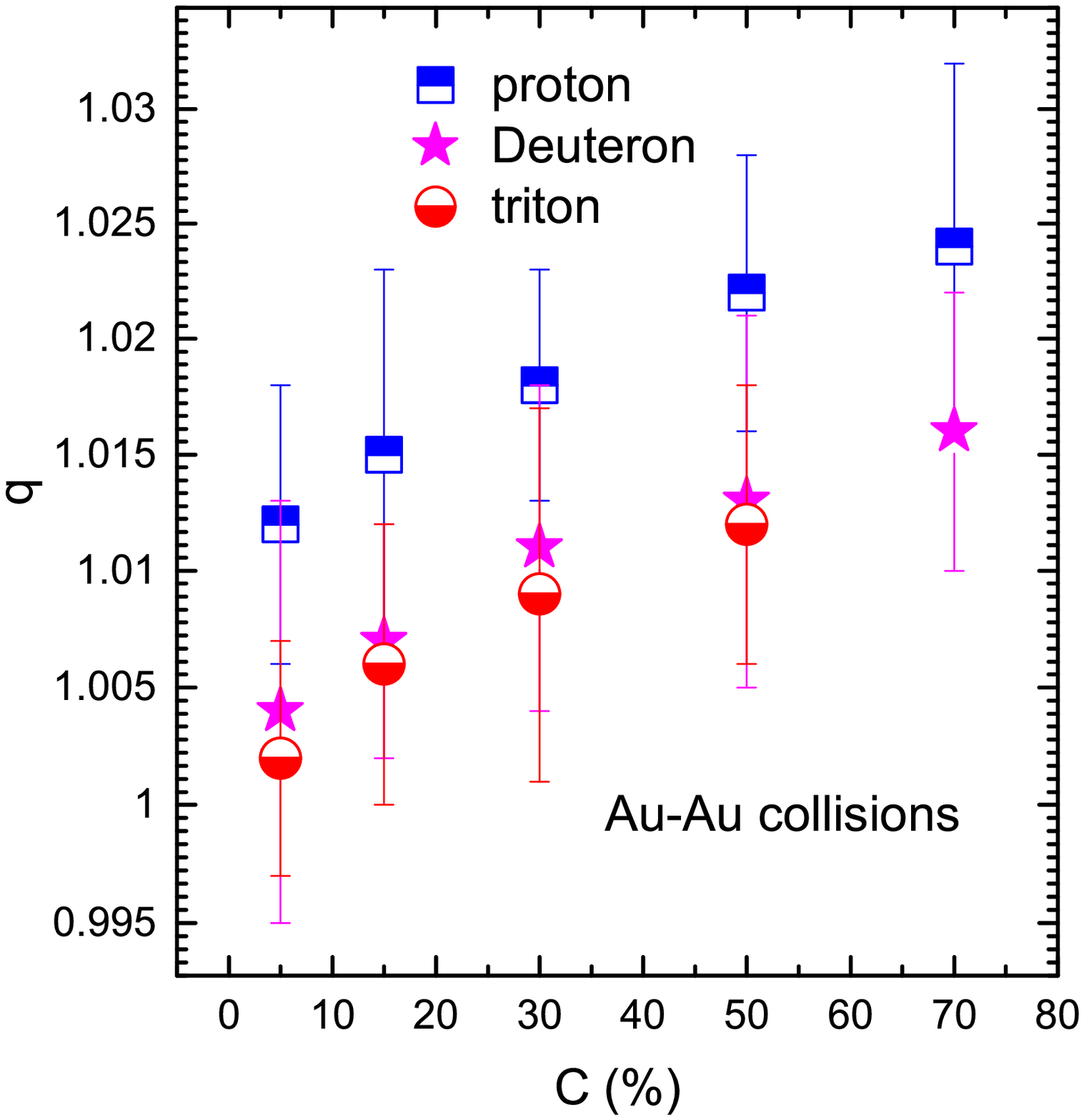}
\end{center}
Fig. 7 Dependence of $q$ on centrality.
\end{figure*}

Fig. 7 shows the dependence of $q$ on centrality. The parameter
$q$ is smaller in central collisions but as we go from central to
peripheral collisions it is going to increase. The parameter is
varying with mass of the particle, larger $q$ is observed for
light particles. The interacting system stays at equilibrium
because $q$ is being very close to 1.
\begin{figure*}[htb!]
\begin{center}
\hskip-0.153cm
\includegraphics[width=15cm]{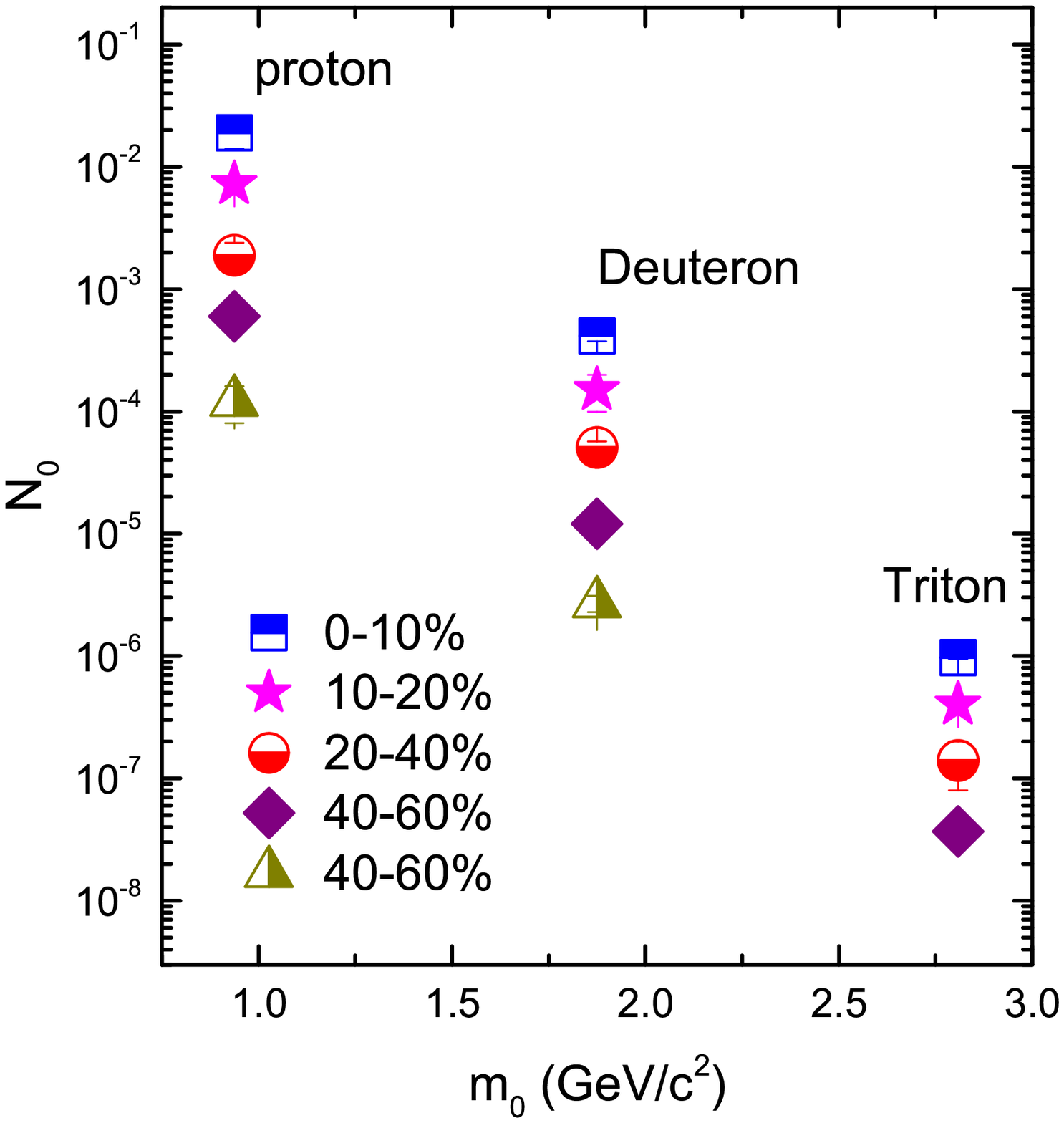}
\end{center}
Fig. 8. Dependence of $N_0$ on mass and centrality.
\end{figure*}

In  Fig 8 the parameter $N_0$ decrease with the mass of the particle. $N_0$ basically shows the multiplicity and it is larger in central collision which decrease towards periphery
\\
{\bf Further discussion 3.2}

 The study of $p_T$ spectra of the particles may give some fruitful information about effective temperature ($T_{eff}$), initial temperature ($T_i$), thermal/kinetic freezeout temperature ($T_0$), thermal freezeout volume ($V$) of the interacting system, and transverse flow velocity ($\beta_T$) of the final sate particles. We use the fitting method to extract these information by using different models and distributions. In the present work, the blast wave model with Tsallis statistics is used.

The structure of transverse momentum ($p_T$) spectra of charged particles generated in high energy heavy ion collisions is very complex. It is not enough to use only one probability density function to describe the $p_T$ spectra, though this function can be of various forms. Particularly, in case when the maximum $p_T$ reaches to 100 GeV at LHC collisions [41]. Several $p_T$ regions are observed by the model analysis [42], including the first region $p_T$$<$ 4-6 Gev/c, 4-6 Gev/c$<$$p_T$$<$17-20 GeV/c and the third region with $p_T$$>$17-20 GeV/c. The boundaries of different $p_T$ regions at the RHIC are slightly lower. It is expected that different $p_T$ regions  correspond to different interacting mechanisms. Even for the same $p_T$ region, there are different explanations due to different model methods and microcosmic pictures.

According to [42], different whole features of fragmentation and hadronization of partons through the string dynamics corresponds to different $p_T$ regions. The effects and changes by the medium take part in the main role in the first $p_T$ region, while it has weak appearance in second $p_T$ region. At the same time, in the third region, the nuclear transparency results in negligible impact of the medium. From number of strings point of view, the maximum number of strings are in second $p_T$ region that results in fusion and creation of strings and collective behavior of partons. The second $p_T$ region is proposed as a possible area of Quark-Gluon Plasma (QGP) through the string fusion. While due to direct hadronization of the low energy strings into mesons [42], the first $p_T$ region has the minimum number of strings and maximum number of hadrons. In some cases there may be the contribution region ($p_T$ $<$ 0.2-0.3 GeV/c) of very soft process which is due to resonant production of charged particles e:g pions, and this region is considered as the fourth $p_T$ region. Different components in a unified superposition can describe the four $p_T$ regions. We have two methods in order to structure the unified superposition. First method is the common method of overlapping of the contribution regions of various components, however the second method is the Hagedorn model [38] which exclude this overlapping. If the contribution of hard component in the first method is be neglected in low $p_T$ region due to its small value, the first method can be changed into second method. Indeed the contribution to $T_0$ and $\beta_T$ is less for the hard component. If the spectra in low $p_T$ region is analyzed to extract only $T_0$ and $\beta_T$, then we can give up the second part of eqs.(3) and (4). That is, $f_S(p_T)$ can be used directly from eqs. (1) which also includes the contribution of very-soft component that comes from resonance decays if available in the data. In the present work, the contribution of hard component in low $p_T$ region if available is included in the extraction of $T_0$ and $\beta_T$ which may cause a slight increase in $T_0$ and/or $\beta_T$ but the relative increase can be neglected due to small values [43]. In the present work we only use eqn. (1), which means that the fraction of hard component is zero in low $p_T$ region. But we also show eqs.(3) and (4) to show a method for further analysis if necessary.
\\
 {\section{Conclusions}}
 The main observations and conclusions are summarized here.

 a) The transverse momentum spectra of proton ($p$), deuteron ($d$) and triton ($t$) are analyzed by the blast wave model with Tsallis statistics and the bulk properties in terms of the kinetic freeze out temperature, transverse flow velocity and kinetic freeze out volume are extracted.

 b) The kinetic freeze out temperature ($T_0$) is observed to increase from central to peripheral collisions. However the transverse flow velocity and freeze out volume is decreasing from central to peripheral collision.

 c) The entropy index ($q$) increasing with from centrality while the parameter $N_0$ is decreasing with centrality.

 d) The kinetic freeze out temperature, transverse flow velocity and freeze out volume decrease with the increasing mass of the particle. Therefore mass differential kinetic freeze out scenario and volume differential freeze out scenario is observed.

 e) both the entropy index ($q$) and the parameter $N_0$ decreasing the mass of particle.

   )
\\
\\

{\bf Data availability}

The data used to support the findings of this study are included
within the article and are cited at relevant places within the
text as references.
\\
\\
{\bf Compliance with Ethical Standards}

The authors declare that they are in compliance with ethical
standards regarding the content of this paper.
\\
\\
{\bf Acknowledgements}

The authors would like to thank support from the National Natural Science
Foundation of China (Grant Nos. 11875052, 11575190, and 11135011).
\\
\\

\end{multicols}

{\small
\begin{thebibliography}{99}
\setlength{\itemsep}{-1pt}
\bibitem{1}
J.~Cleymans, H.~Oeschler, K.~Redlich and S.~Wheaton,
Phys. Rev. C \textbf{73} (2006), 034905
doi:10.1103/PhysRevC.73.034905
[arXiv:hep-ph/0511094 [hep-ph]].

\bibitem{2}
Itoh N 1970
A. Andronic, P. Braun-Munzinger and J. Stachel, 
Nucl. Phys. A, Vol. 834, pp 240, 2010

\bibitem{3}
A. Andronic, P. Braun-Munzinger and J. Stachel, Acta
Phys. Polon. B, Vol. 40, pp 1005-1012, 2009.

\bibitem{4}
Y. Hama, F.S. Navarra, Z. Phys. C, Vol. 53, pp
501C506, 1992.

\bibitem{5}
E. Schnedermann, J. Sollfrank, U. Heinz, Phys. Rev. C,
Vol. 48, pp 2462C2475, 1993.

\bibitem{6}
B.I. Abelev, et al. [STAR Collaboration]. Phys. Rev. C,
Vol. 79, article 034909, 2009.

\bibitem{7}
B.I. Abelev, et al. [STAR Collaboration]. Phys. Rev. C,
Vol. 81, article 024911, 2010.

\bibitem{8}
Z. Tang, Y. Xu, L. Ruan, G. van Buren, F. Wang and
Z. Xu, Phys. Rev. C, Vol. 79, article 051901, 2009.

\bibitem{9}
Z. Tang, L. Yi, L. Ruan, M. Shao, H. Chen, C. Li,
B. Mohanty, P. Sorensen, A. Tang and Z. Xu,Chin.
Phys. Lett. Vol. 30, article 031201, 2013.

\bibitem{10}
K. Jiang, Y. Zhu, W. Liu, H. Chen, C. Li, L. Ruan,
Z. Tang, Z. Xu and Z. Xu,Phys. Rev. C, Vol. 91, article 024910, 2015.

\bibitem{11}
S. Takeuchi, K. Murase, T. Hirano, P. Huovinen and
Y. Nara, Phys. Rev. C, Vol. 92, article 044907, 2015.

\bibitem{12}
H. Heiselberg, A.M. Levy, Phys. Rev. C,Vol. 59, pp
2716C2727, 1999.

\bibitem{13}
U.W. Heinz, Concepts of heavy-ion physics. Lec-
ture notes for lectures presented at the 2nd CERNC
Latin-American school of high-energy physics, 1C14
June 2003, San Miguel Regla, Mexico. arXiv:hep-
ph/0407360.

\bibitem{14}
R. Russo, PhD Thesis, Universita degli Studi di Torino,
Italy. arXiv:1511.04380 [nucl-ex].

\bibitem{15}
H.-R. Wei, Fu.Hu. Liu, R.A. Lacey, Eur. Phys. J. A,
Vol. 52, article 102, 2016.

\bibitem{16}
H. L. Lao, H. R. Wei, Fu-Hu. Liu and R. A. Lacey, Eur.
Phys. J. A, Vol. 52, article 203, 2016.

\bibitem{17}
H. R. Wei, Fu-Hu-Liu and R. A. Lacey,. Disentangling
random thermal motion of particles and collective ex-
pansion of source from transverse momentum spectra
in high energy collisions. J. Phys. G, Vol. 43, article
125102, 2016.

\bibitem{18}
M.Waqas, Fu-Hu-Liu, S. Fakhraddin and M. A. Rahim,
J. Phys. Vol.93, pp 1329-1343, 2019.

\bibitem{19}
M. Waqas and Fu-Hu-Liu, [arXiv:1806.05863 [hep-ph]].

\bibitem{20}
M. Waqas and Fu-Hu-Liu, Eur. Phys. J. Plus Vol. 135,
article 147, 2020.

\bibitem{21}
 Q. Wang and Fu-Hu-Liu, Int. J. Theor. Phys. Vol. 58,
pp 4119-4138, 2019.

\bibitem{22}
L. Kumar [STAR], Nucl. Phys. A Vol. 931, pp 1114-
1119, 2014.

\bibitem{23}
L. Adamczyk et al. [STAR], Phys. Rev. C, Vol. 96, ar-
ticle 044904, 2017.

\bibitem{24}
L. Adamczyk et al. (STAR Collaboration), Phys. Rev.
C Vol. 94, article 034908, 2016.

\bibitem{25}
J. Adam et al. (ALICE Collaboration), Phys. Rev. C,
Vol. 93, article 024917, 2016.

\bibitem{26}
J. Chen, D. Keane, Y. G. Ma, A. Tang, and Z. Xu,
Phys. Rept. Vol 760, articcle 1, 2018.

\bibitem{27}
H. H. Gutbrod, A. Sandoval, P. J. Johansen, A. M.
Poskanzer, J. Gosset, W. G. Meyer, G. D. Westfall, and
R. Stock, Phys. Rev. Lett. Vol. 37, pp 667C670, 1976.

\bibitem{28}
Rudiger Scheibl and Ulrich Heinz, Phys. Rev. C, Vol.
59, pp 1585C1602, 1999.

\bibitem{29}
W. J. Llope et al., Phys. Rev. C, Vol. 52, pp 2004C2012,
1995.

\bibitem{30}
H. Sato and K. Yazaki, Phys. Lett. B Vol. 98, pp 153 C
157, 1981.

\bibitem{31}
Kai-Jia Sun, Lie-Wen Chen, Che Ming Ko, and Zhang-
bu Xu, Phys. Lett. B, Vol. 774, pp 103 C 107, 2017.

\bibitem{32}
S. T. Butler and C. A. Pearson, Phys. Rev. Vol. 129,
article 836, 1963.

\bibitem{33}
M. Waqas, F. H. Liu and Z. Wazir, Adv. High Energy
Phys. Vol. 2020, article 8198126, 2020.

\bibitem{34}
M. Waqas, Fu-Hu-Liu, L. L. Li and H. M. Alfanda, Nu-
cl. Sci. Tech. Vol. 31, article 109, 2020.

\bibitem{35}
R. Odorico, Phys. Lett. B, Vol. 118, pp 151C154, 1982.

\bibitem{36}
Arnison, G.; et al. [UA1 Collaboration]. Phys. Lett. B
Vol. 118, article 167C172, 1982.

\bibitem{37}
Mizoguchi, T.; Biyajima, M.; Suzuki, N. Int. J. Mod.
Phys. A, Vol. 32, article 1750057, 2017.

\bibitem{38}
Hagedorn, R. Multiplicities, Riv. Nuovo Cimento, Vol.
6, article 1, 1983.

\bibitem{39}
Abelev, B.; et al. [ALICE Collaboration].Eur. Phys. J.
C Vol. 75, article 1, 2015.

\bibitem{40}
Dingwei Zhang (STAR Collaboration),GeVXXVIIIth
International Conference on Ultrarelativistic Nucleus-
Nucleus Collisions (Quark Matter 2019), Nuclear
Physics A Vol. 00, pp 1C4, 2020.

\bibitem{41}
S Chatrchyan et al. [CMS Collaboration] Eur. Phys. J.
C vol 72, 1945, 2012

\bibitem{42}
M K Suleymanov Int. J. Mod. Phys. E Vol, 27, 1850008,
2018

\bibitem{43}
H-L Lao, F-H Liu, B-C Li, M-Y Duan and R A Lacey
Nucl. Sci. Tech. 29 164 (2018)

\
\end{thebibliography}}
\end{document}